\documentclass[twocolumn,prd,aps,superscriptaddress
,showpacs,
nofootinbib,preprintnumbers,floats,floatfix]{revtex4}

\usepackage{color}
\usepackage{amsmath}
\usepackage{amssymb}
\usepackage{latexsym}
\usepackage{graphicx}
\usepackage{hyperref}
\usepackage{bm}

\newcommand{\fnl}{\ensuremath{f_{\mathrm{NL}}}}

\newcommand{\mpl}{m_{\rm Pl}}

\begin{document}

\title{Comprehensive analysis of the simplest curvaton model}
\author{Christian T.~Byrnes}
\affiliation{Astronomy Centre, University of Sussex, Falmer,  
Brighton BN1 9QH, UK}
\author{Marina Cort\^es} 
\affiliation{Institute for Astronomy, University of Edinburgh, Blackford Hill, Edinburgh EH9 3HJ, UK} 
\affiliation{Centro de Astronomia e Astrof\'isica da Universidade de Lisboa, Faculdade de Ci\^encias, Edif\'icio C8, Campo Grande, 1769-016 Lisboa, Portugal}
\affiliation{Perimeter Institute for Theoretical Physics
31 Caroline Street North, Waterloo, Ontario N2J 2Y5, Canada}
\author{Andrew R.~Liddle} 
\affiliation{Institute for Astronomy, University of Edinburgh, Blackford Hill, Edinburgh EH9 3HJ, UK} 
\date{\today}
\begin{abstract}
We carry out a comprehensive analysis of the simplest curvaton model, which is based on two non-interacting massive fields. Our analysis encompasses cases where the inflaton and curvaton both contribute to observable perturbations, and where the curvaton itself drives a  second period of inflation. We consider both power spectrum and non-Gaussianity observables, and focus on presenting constraints in model parameter space. The fully curvaton-dominated regime is in some tension with observational data,  while an admixture of inflaton-generated perturbations improves the fit. The inflating curvaton regime mimics the predictions of Nflation. Some parts of parameter space permitted by power spectrum data are excluded by non-Gaussianity constraints. The recent BICEP2 results [1] require that the inflaton perturbations provide a significant fraction of the total perturbation, ruling out the usual curvaton scenario in which the inflaton perturbations are negligible, though not the admixture regime where both inflaton and curvaton contribute to the spectrum.
\end{abstract}

\pacs{98.80.Cq}

\maketitle

\section{Introduction}

While observational results, including recent ones from the {\it Planck} mission \cite{planckI,planckXXII} and from BICEP2 \cite{bicep2}, continue to strongly support inflation as the origin of cosmic structure, it remains an open issue whether the observed perturbations arise from fluctuations in the field driving inflation or from a different degree of freedom. A particular example of the latter class is the curvaton model \cite{seminal}, and there have been several reports on the status of those models  in the light of {\it Planck} satellite results \cite{Enq-Tak,kobayashi,ellis,higgs,tarrant}. In this work we carry out an analysis of the simplest curvaton model \cite{BL}, aiming at an exhaustive study of parameter space while minimizing the set of usual assumptions. 

Our analysis is principally analytical. We consider wide regimes of relative inflaton/curvaton contribution to the curvature perturbation and energy densities. We extend the existing literature in several directions. We impose simultaneous constraints from the full set of observables in the model parameter space. We provide a detailed modelling of the number of $e$-foldings corresponding to observable scales --- the so-called `pivot' scale \cite{CLM} --- and allow it to respond to the change in inflationary energy scale in different parts of parameter space. We include the effect of the curvaton mass on perturbations generated via the curvaton, and we consider the region of parameter space where the curvaton may itself drive a second period of inflation. 

After ensuring that the accurately-observed perturbation amplitude is reproduced, and once a reheating model is selected, the model reduces to three parameters which can be taken as the masses of the inflaton and curvaton and the value of the curvaton when the pivot scale crosses the horizon. Each observable depends on at most two of these, but in different combinations. 
Allowing for arbitrary decay times of the both fields extends the parameter space, which we parametrize by the number of matter-like $e$-foldings. This is due to the pressureless equation of state while a field oscillates in a quadratic potential. We do not make assumptions for the time of curvaton decay, nor the relative size of field masses. Extending the analysis of Ref.~\cite{Enq-Tak}, we show that the curvaton mass can be comparable but not significantly greater than the inflaton mass.

\section{Curvaton models}

\label{s:2}

In early papers \cite{seminal}, the curvaton was assumed to be a second, light, scalar field present during inflation which
\begin{enumerate}
\item has a subdominant energy density compared to the inflaton's, while the inflaton drives inflation.
\item is long lived (i.e.~it decays later than the inflaton).
\item\label{curvaton-limit} generates the entire primordial curvature perturbation.
\end{enumerate}
In common with many other papers, we will abandon assumption 3 to include the mixed inflaton--curvaton scenario. We later discuss the case where the curvaton itself drives a short period of inflation \cite{Langlois:2004nn,Dimopoulos:2011gb, moroi_takahashi_toyoda,Senoguz:2012iz}, which is permitted by the above assumptions though this possibility is often ignored.

Throughout we denote the inflaton field by $\phi$, defined as the field which dominates the energy density when observable scales first cross outside the horizon, and the curvaton field by $\sigma$ (though in some parameter regimes the curvaton can contribute a late-stage era of inflation). 
We focus on the simplest curvaton model \cite{BL}, featuring two massive non-interacting fields with potential
\begin{equation}
V(\phi,\sigma) = \frac{1}{2}m_\phi^2 \phi^2+ \frac{1}{2}m_\sigma^2 \sigma^2. 
\end{equation}
The number of $e$-foldings of inflation from field values $\phi$ and $\sigma$ is given by 
\begin{equation}
N  = 2\pi \frac{\phi^2 + \sigma^2}{m_{\rm Pl}^2},
\end{equation}
where $m_{\rm Pl}$ is the (non-reduced) Planck mass and we have neglected the small contributions from the field values at the end of inflation.

Like the authors of Ref.~\cite{Enq-Tak}, we consider the full range from negligible to full curvaton contribution to the total power spectrum, given by:
\begin{equation}\label{ps}
P^{\rm total}_{\zeta}= P^{\phi}_{\zeta}+P^{\sigma}_{\zeta}\,.
\end{equation}
We can parametrize the inflaton contribution to the total power spectrum as
\begin{equation}\label{pzeta}
P^{\phi}_{\zeta} = \frac{m_{\phi}^2}{m_{\rm single}^2} P^{\rm total}_{\zeta}.
\end{equation}
Here $m_{\rm single}$ is the mass that the inflaton would need if it were to give the correct amplitude of perturbations in the single-field case; in a scenario where both field contribute this is an upper limit to the actual inflaton mass $m_\phi$.  It is determined by
\begin{eqnarray}
P_{\rm single} &=& \frac{8 V_{\rm single}}{3 m_{\rm Pl}^4 \epsilon_{\rm single}}\bigg|_{*}\\
&=& \frac{4 m_{\rm single}^2 \phi_{\rm single}^2}{3 m_{\rm Pl}^4} \,2N\bigg|_{*}
\end{eqnarray}
where * refers to the parameter value when observable scales crossed the Hubble radius during inflation, $V_{\rm single}= m_{\rm single}^2 \phi_{\rm single}^2/2$,  and 
\begin{equation}
\epsilon_{\rm single} \equiv \frac{m_{\rm Pl}^2}{16 \pi} \,  \left( \frac{V'}{V}\right)^2=\frac{1}{2N_*}
\end{equation} 
in the single-field model. Taking the observed amplitude as \cite{WMAP9, planckXVI}
\begin{equation} \label{ampNorm}
P_\zeta^{\rm obs} \sim 2.2\times 10^{-9}\,,
\end{equation}
we obtain 
\begin{equation}
\frac{m_{\rm single}^2}{m_{\rm Pl}^2}=5.2\times10^{-9}\frac{1}{N_*^2}.
\end{equation}
The ratio $m_{\phi}^2/m_{\rm single}^2$ will appear throughout in our expressions as a measure of the relative contribution of the inflaton to the power spectrum in each model.

The curvaton contribution to the power spectrum is determined by the ratio of curvaton to background energy density at the time the curvaton decays into the thermal bath:
\begin{equation}\label{rdec}
r_{\rm dec} \equiv \frac{3 \rho_\sigma}{4 \rho_\gamma+3 \rho_\sigma}\bigg|_{\rm decay}
\end{equation}
where we assumed that the inflaton has fully decayed into radiation before the curvaton decays. 

Equation~(\ref{rdec}) is defined so as to provide a unified expression for the curvaton perturbation in the regimes of radiation and curvaton domination at the time of decay, which is \cite{Lyth:2002my}
\begin{equation}\label{Psigma}
P_\zeta^{\sigma} = \frac{r_{\rm dec}^2}{9 \pi^2}\frac{H_*^2}{\sigma_*^2}\,.
\end{equation}
We use the normalization amplitude Eq.~(\ref{ampNorm}) to fix the ratio $r_{\rm dec}^2 H_*^2/\sigma_*^2$ and obtain 
\begin{equation}\label{rdec-our}
r_{\rm dec}^2  = 5.9 \times 10^{-7}  \left(1-\frac{m_{\phi}^2}{m_{\rm single}^2}\right)  \frac{\sigma_{*}^2}{2 m_{\phi}^2 N_*}
\end{equation}
where henceforth $N_*$ is the $e$-foldings number at which the {\it Planck} normalization scale $0.05\, {\rm Mpc}^{-1}$ crosses the Hubble radius during inflation. Evaluating Eq.~(\ref{rdec}) requires knowledge of the full curvaton evolution, but in practice we will only use $r_{\rm dec}$ via Eq.~(\ref{rdec-our}) as a constraint on model parameters by requiring that it takes the physically realisable values $0<r_{\rm dec}<1$, in Sec.~\ref{sec:NonG} we will see that the lower bound is tightened by the constraint on local $\fnl$. We may apply this constraint even if the curvaton rolls significantly during inflation, i.e.~if $m_\sigma\simeq m_\phi$ \cite{Sasaki:2006kq}.

\section{Parametrization of the number of $e$-foldings}\label{nefold}

To impose accurate constraints we need to identify the correct number of $e$-foldings corresponding to the pivot scale at which observables are evaluated. The number of $e$-foldings that occurred after exit of the current Hubble scale is given by \cite{LiddleLeach}
\begin{equation}\label{Nhor}
N_{\rm hor}=63+\frac{1}{4}\ln \epsilon+ \frac{1}{4}\ln\frac{V_{\rm hor}}{\rho_{\rm end}}+\frac{1}{12}\ln\frac{\rho_{\rm reh}}{\rho_{\rm end}}\,,
\end{equation}
where all quantities are as in single-field models. We parametrize observables as a function of the number of $e$-folds before the end of inflation, when the corresponding scale left the horizon, and so Eq.~(\ref{Nhor}) gets a correction to account for the difference between the Hubble length for which it holds, and the observable scale we measure at. For the {\it Planck} pivot $k=0.05 \, {\rm  Mpc}^{-1}$ we get, 
\begin{equation}
N_* \cong N_{\rm hor}-5\,.
\end{equation}
We also parametrize the total amount of reheating $e$-foldings, given by the last term of Eq.~(\ref{Nhor}), as $N_{\rm matter}$ which includes the reheating of both the inflaton and the curvaton, obtaining 
\begin{equation}
\frac{1}{12}\ln\frac{\rho_{\rm reh}}{\rho_{\rm end}} =- \frac{1}{4}N_{\rm matter}\,.
\end{equation}
For the quadratic inflaton, the middle two terms in Eq.~(\ref{Nhor}) combine into a term which measures the inflaton mass relative to the single-field limit (i.e.\ those terms cancel in the single-field case), giving
\begin{equation}\label{Nstar}
N_* =  58+\frac{1}{2} \ln{\frac{m_{\phi}}{m_{\rm single}}-\frac{1}{4} N_{\rm matter}}.
\end{equation}
We note that this relation sets a firm upper limit of $58$ as the number of $e$-foldings corresponding to the {\it Planck} pivot scale. This seems to contradict some values quoted in Ref.~\cite{ellis}. 

Instead of parametrizing the uncertainty in associating a pivot scale by $\Gamma_\phi$ or $N_*$, we take the number of matter-like $e$-foldings from the end of inflation to the decay of the curvaton, denoted $N_{\rm matter}$. This parameter has quite a wide plausible range, for instance $N_{\rm matter} = 0$ implies instant reheating and no curvaton domination, while positive values allow for both the period of reheating after inflation and any subsequent period of curvaton domination.  $N_{\rm matter} > 16$ would ensure reheating at less than $10^{11} {\rm GeV}$ to evade overproduction of gravitinos, while $N_{\rm matter} \lesssim 40$ is necessary to ensure reheating before electro-weak symmetry breaking.  For our main results we take a central value of $N_{\rm matter} = 20$, which in the single-field case gives $N_* = 53$ with a plausible modelling uncertainty of 5 in either direction, in keeping with usual estimates.

If the BICEP2 detection of $r$ is confirmed, then $m_\phi$ cannot be significantly below the single-field value, and the relevant term in Eq.~(\ref{Nstar}) is negligible. If we do not impose a lower bound on $r$, then this term may become large, but reducing the energy scale of inflation also reduces the maximum permitted number of matter $e$-foldings, partially cancelling this effect. In the most extreme case, when inflation ends as late as possible and the curvaton decays shortly before nucleosynthesis, we estimate that it is possible to reduce the $e$-foldings further to $N_*\simeq44$.

\section{Observables and model constraints}
\subsection{Linear power spectrum}

We can now make predictions for model observables: the spectral index $n_{\rm S}$, tensor-to-scalar ratio $r$, and non-Gaussianity parameter $\fnl$. The slow-roll parameters are defined by
\begin{equation}
\epsilon=-\frac{\dot{H}}{H^2}\simeq \frac12 \left(\frac{V_{\phi}}{3H^2}\right)^2, \;\;\eta_\phi = \frac{V_{\phi \phi}}{3H^2},\;\;\eta_\sigma = \frac{V_{\sigma \sigma}}{3 H^2}.
\end{equation}
If there is a single period of inflation, $\epsilon \simeq \eta_\phi \simeq 2/N_*$.

The predicted deviation from scale invariance is \cite{Wands:2002bn},
\begin{equation}\label{ns}
n_{\rm S}-1= \left(1-\frac{m_{\phi}^2}{m_{\rm single}^2}\right) (-2\epsilon+2\eta_{\sigma}) +\frac{m_{\phi}^2}{m_{\rm single}^2}\left(-6\epsilon+2 \eta_{\phi}\right).
\end{equation}
This expression allows for ready distinction between the contribution of the inflaton versus curvaton for each model spectrum, as a function of the inflaton mass, $m_{\phi}$. It reduces to each of these two regimes for $m_{\phi} \sim m_{\rm single}$ and $m_{\phi} \ll m_{\rm single}$ respectively. We further note that $n_{\rm S}$ has no dependence on $\sigma_*$. Lastly, Eq.~(\ref{ns}) has an implicit dependence on the number of $e$-foldings and the chosen pivot scale, which we presented in Section~\ref{nefold}.

The tensor-to-scalar ratio $r$ is readily obtainable by
\begin{equation}\label{r_tens}
r =16\, \epsilon \frac{m_{\phi}^2}{m_{\rm single}^2} \,,\\
\end{equation}
also parametrized by the relative contribution of the inflaton mass, and without further dependence on curvaton parameters. As $m_\phi$ is varied from zero to $m_{\rm single}$, the prediction in the $n_{\rm S}$--$r$ plane interpolates linearly between the curvaton-dominated and inflaton-dominated regimes.

\begin{figure}[t!]
\includegraphics[width=0.45 \textwidth]{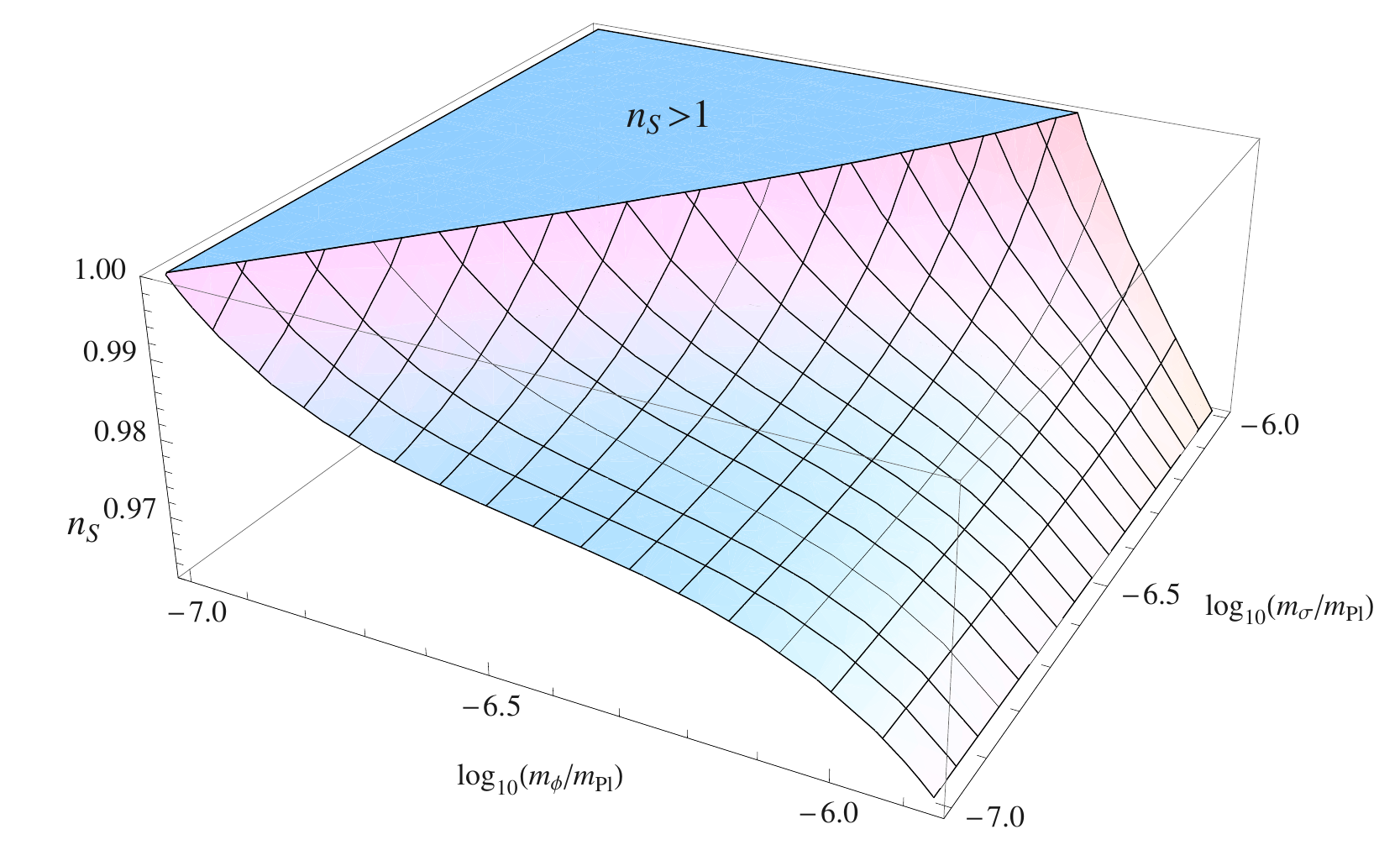}\\
\vspace*{0.5cm}
\includegraphics[width=0.45 \textwidth]{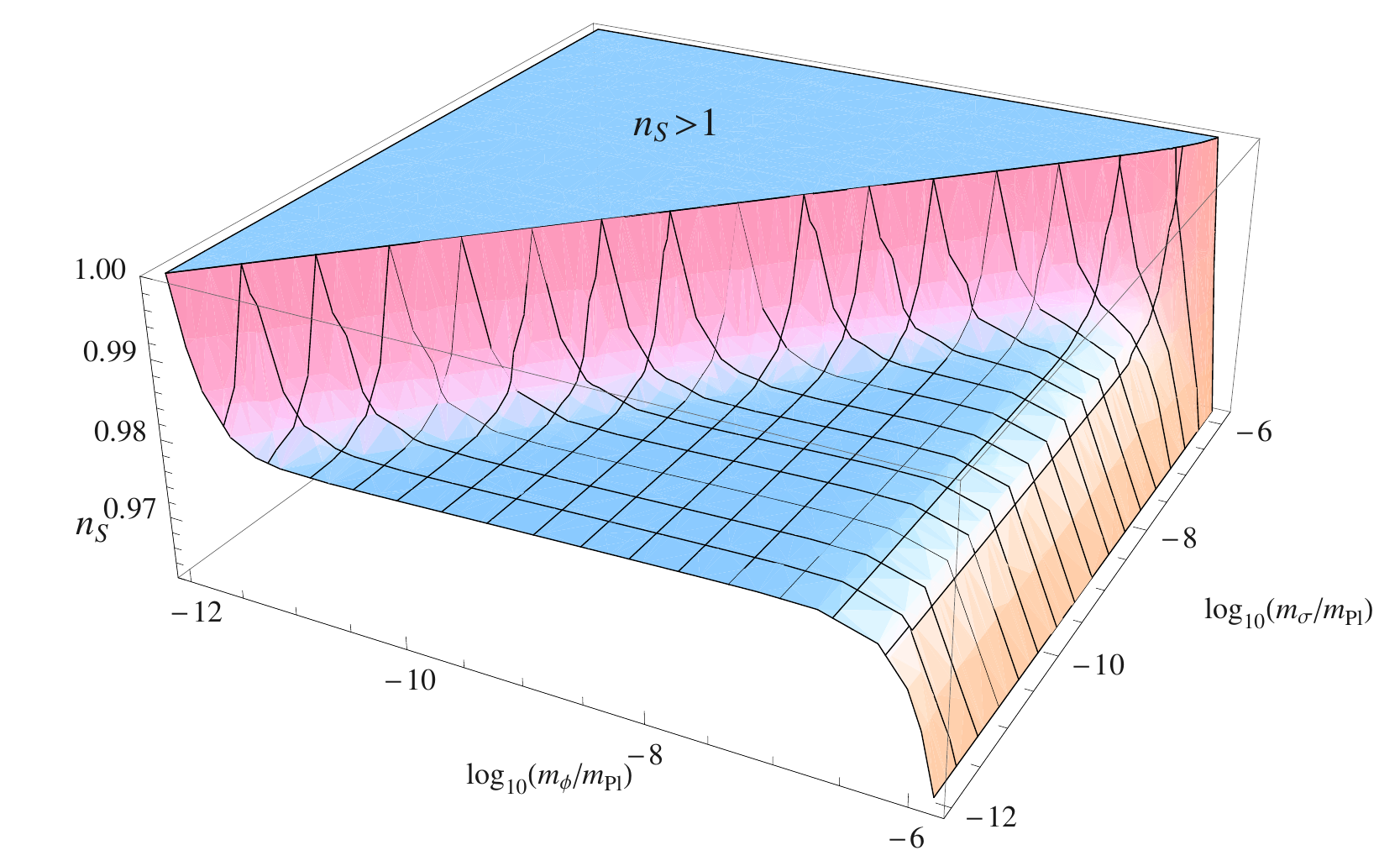}
\caption{The prediction for $n_{\rm S}$ shown for two different ranges of parameters, and cut-off where $n_{\rm S}$ goes above one. This figure takes $N_{\rm matter} = 20$.
}
\label{ns1} 
\end{figure}

In Fig.~\ref{ns1} we show the prediction for $n_{\rm S}$ given by Eq.~(\ref{ns}), across two different ranges of field masses. Given that $n_{\rm S}$ has no dependence on $\sigma_*$, the space of model parameters is two-dimensional. The right edge is the inflaton-dominated regime with $n_{\rm S} \simeq 0.96$, while the large flat area at $n_{\rm S} \simeq 0.98$ is the region which is curvaton-dominated and in which the curvaton has a negligible mass. This is more or less where the 95\% upper limit on $n_{\rm S}$ lies according to data compilations including {\em Planck} results \cite{planckXXII}, and hence whether these models can be considered allowed or excluded at this level is sensitive to precise data compilation choice, to the choice of parameters varied in the cosmological fits, and to the modelling of $N_*$.
For large enough $m_\sigma$ the spectral index rises due to the curvaton mass, crossing $n_{\rm S} = 1$ at $m_\sigma \simeq m_\phi$ (since for this inflaton potential $\epsilon \simeq \eta_\phi$). 

\begin{figure}
\includegraphics[width=0.45 \textwidth]{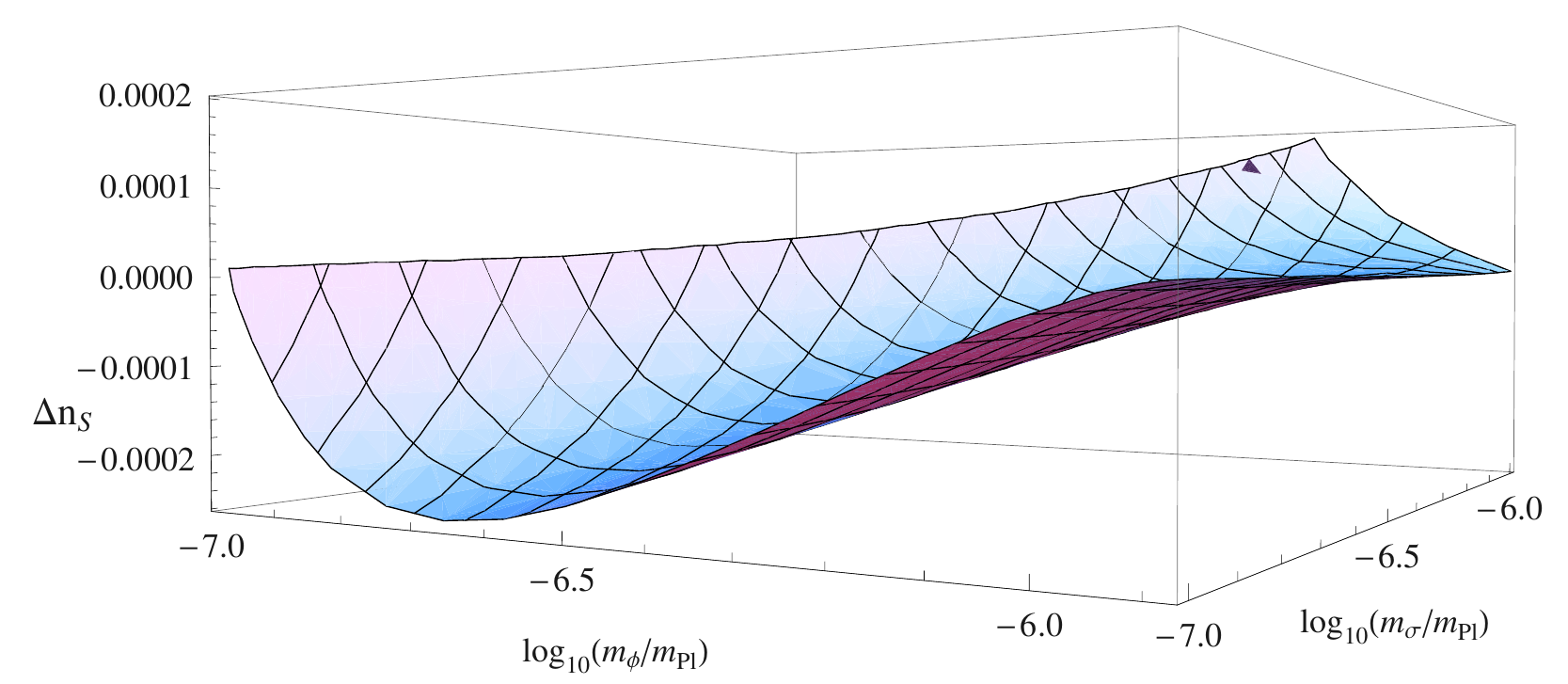}
\caption{The difference in $n_{\rm S}$ arising from taking $N_{\rm matter} = 20$ rather than $N_* = 53$. By design they agree at the inflaton limit.}
\label{ns_diff} 
\end{figure}

Figure \ref{ns_diff} shows the difference between the $n_{\rm S}$ prediction from our method of parametrizing $N_*$ in terms of a fixed value of $N_{\rm matter}$, as opposed to choosing a fixed $N_*$. The differences are not large, but neither are they completely negligible at the current observational precision.

\subsection{Non-Gaussianity}\label{sec:NonG}

The curvaton scenario generates non-Gaussianity with the local shape, parametrized by the usual $\fnl$ parameter whose value is \cite{Lyth:2002my,Sasaki:2006kq,Enq-Tak}.
\begin{equation}\label{fnl}
f_{\rm NL} = \frac{5}{12} \left(1-\frac{m_{\phi}^2}{m_{\rm single}^2}\right)^2 \left(\frac{3}{r_{\rm dec}}-4-2 r_{\rm dec}\right).
\end{equation}
Note that this expression is independent of the curvaton mass, which does not appear in the expression for $r_{\rm dec}$. The non-Gaussianity predictions hence also depend on only two model parameters, but in a plane orthogonal to the two that determine $n_{\rm S}$.

In the limit of $m_\phi\ll m_{\rm single}$ this reduces to the standard curvaton result, while the value is suppressed if the (nearly) Gaussian inflaton perturbations also contribute to the total power spectrum; see Ref.~\cite{Enq-Tak} and Eq.~(\ref{pzeta}). If the perturbations from both fields are important then $\fnl$ will have a slow-roll suppressed scale dependence \cite{Byrnes:2010ft}, while in the curvaton limit where one may neglect the inflaton perturbations, $\fnl$ is a constant. This difference is potentially observable \cite{Sefusatti:2009xu,Becker:2012je}. However if the curvaton has self-interactions, $\fnl$ may be strongly scale dependent even in the curvaton limit \cite{Byrnes:2011gh,Kobayashi:2012ba}.

\begin{figure}[t]
\includegraphics[width=0.45 \textwidth]{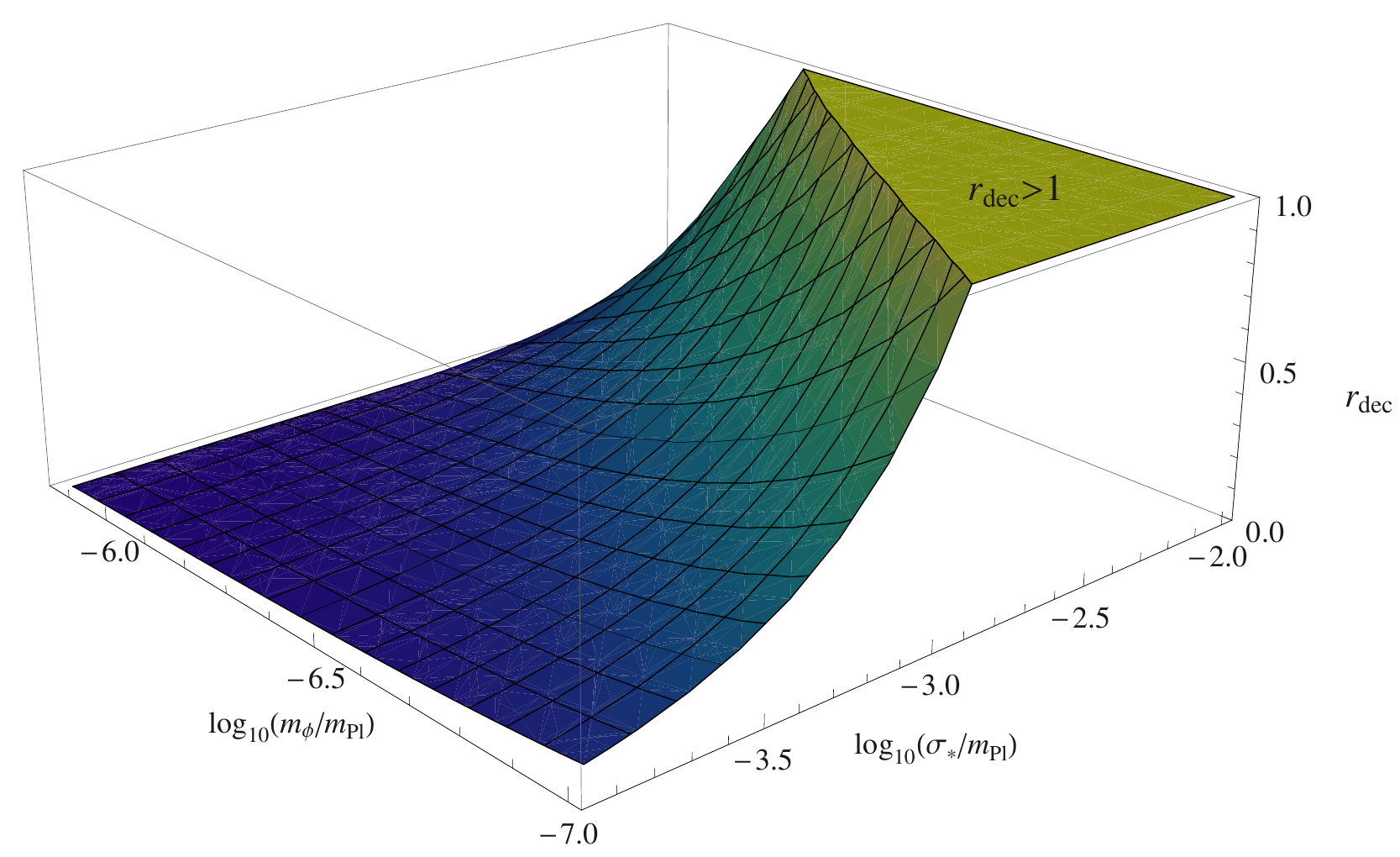}
\caption{Requiring the fraction of curvaton energy density at decay to be physical ($0<r_{\rm dec}<1$) excludes the region of small inflaton mass and large initial curvaton field.}
\label{rdec_physical} 
\end{figure}

\begin{figure}[t]
\includegraphics[width=0.45 \textwidth]{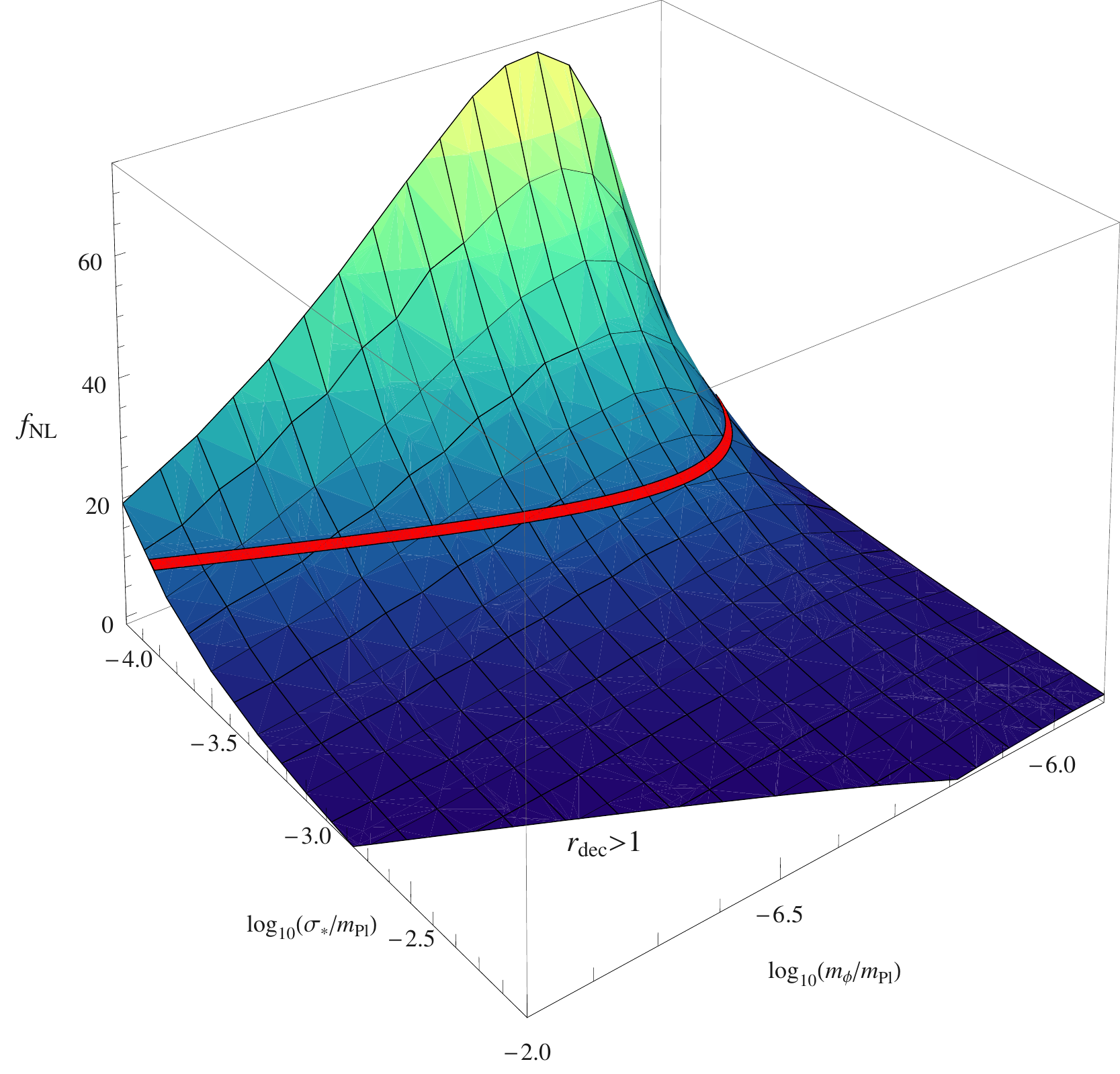}\\
\caption{Values of $\fnl$, showing also the region cut-off by the physical requirement $r_{\rm dec} \le 1$ as in Fig.~\ref{rdec_physical}. The contour corresponds to the 95\% confidence upper limit from {\it Planck}.}
\label{fnl_no_constraint} 
\end{figure}

In Fig.~\ref{rdec_physical} we show the region of parameter space required by $0<r_{\rm dec}<1$. Viable models do not exist outside this region as it is impossible to generate a power spectrum of sufficient amplitude. In Fig.~\ref{fnl_no_constraint} we plot $\fnl$, given by Eq.~(\ref{fnl}), with the unphysical region $r_{\rm dec}>1$ cut off. The red contour marks the 95\% confidence upper limit on $\fnl$ from {\it Planck} \cite{Ade:2013ydc}; the parameter region above this line is excluded.

Probably the only way to rule out the quadratic curvaton model entirely, and independently of the inflationary potential (which may be tuned in order to match any observed value of $n_s$ and $r$), is to detect non-Gaussianity of the local shape satisfying $\fnl < -5/4$. This would even rule out models with two quadratic curvatons \cite{Assadullahi:2007uw}.
\begin{figure*}[t]
\begin{center}
\includegraphics[width=0.5 \textwidth]{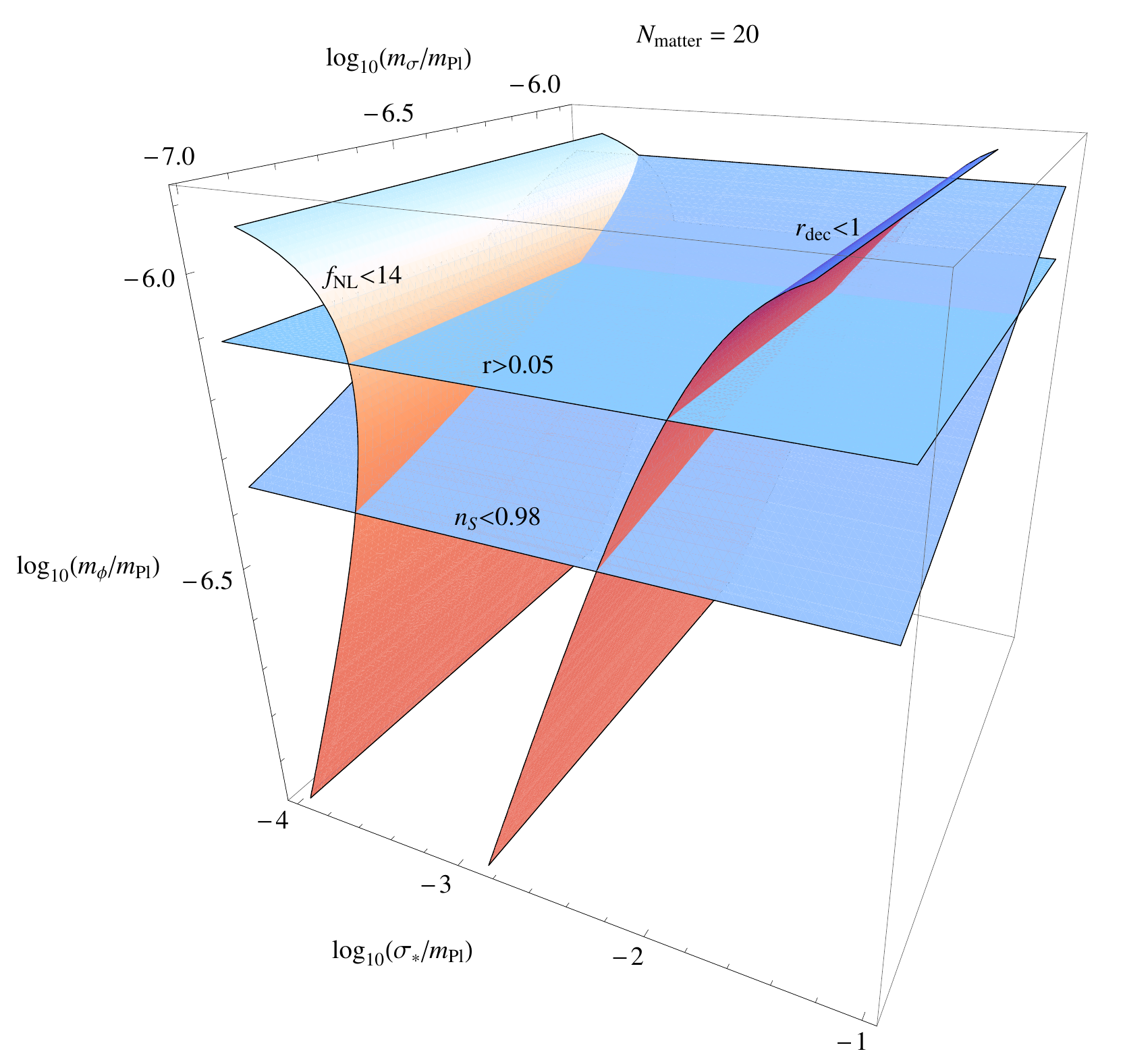}\\ \hspace*{0cm}
\end{center}
\includegraphics[width=0.5 \textwidth]{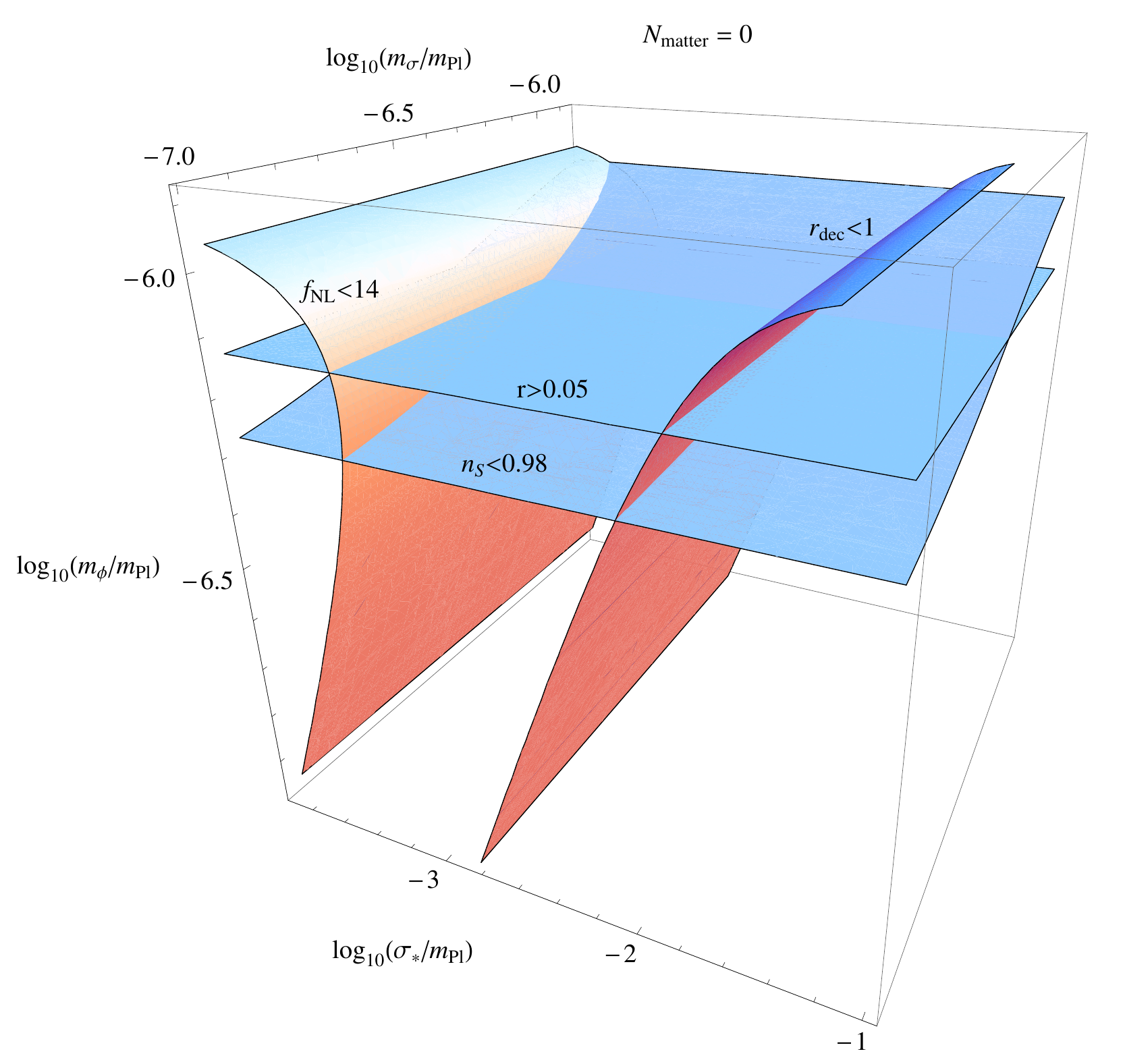}\hspace*{0cm} 
\includegraphics[width=0.5 \textwidth]{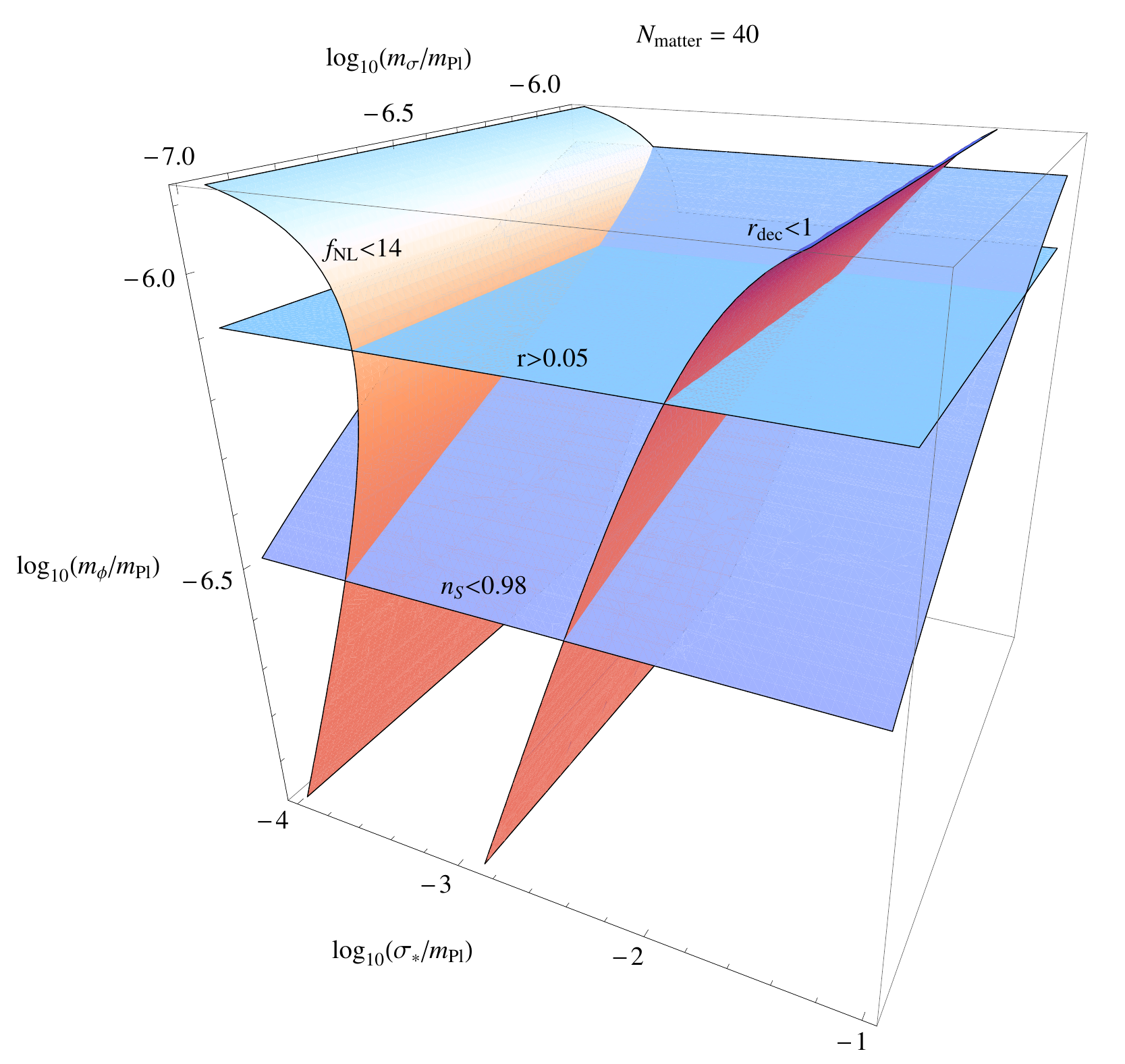} 
\caption{Combined constraints on model space for the fiducial value $N_{\rm matter} = 20$ (top) and extremal values $N_{\rm matter} = 0$ (bottom left) and $N_{\rm matter} = 40$ (bottom right) with limits shown at 95\% confidence for $\fnl=14$,  $n_{\rm S}=0.98$, and we take $r=0.05$ as an indicative lower bound from BICEP2 \cite{bicep2}. Labels are shown to the side of the surface for which they apply.
The allowed region lies between the two mostly vertical planes (non-Gaussianity on the left, physically-allowed value of $r_{\rm dec}$ on the right), and above the plane of constant spectral index, for large values of $m_\sigma$, and above the plane of $r=0.05$ for small curvaton mass. The upper surfaces of the cubes correspond to the inflaton limit $m_{\phi} = m_{\rm single}$ which is the maximum permitted in the model.}
\label{3params} 
\end{figure*}

\subsection{Combined constraints}

Figure~\ref{3params}  shows the allowed region of the full three-dimensional model space,  \{$m_\phi, m_\sigma, \sigma_*$\}, obtained by combining all constraints on $n_{\rm S}$, $\fnl$ and including an indicative lower bound on $r>0.05$ from the new BICEP2 result (the precise value chosen for this makes little difference). We take a fiducial value for $N_{\rm matter} = 20$ in the middle panel and show the effect of varying $N_{\rm matter}$ away from this fiducial to the two extremes of 0 (left) and 40 (right). 
Because the present location of the 95\% limit on $n_{\rm S}$ more or less coincides with the curvaton-dominated regime's prediction, the location of this constraint is extremely sensitive to the exact value chosen, and hence depends on both the choice of data combination used and the assumptions in determining $N_*$. Inclusion of the BICEP2 bound supersedes the $n_{\rm S}$ constraint in the low $m_\sigma$ curvaton regime.

Given these uncertainties, it would thus be premature to say that the strongly curvaton-dominated regime of the model is excluded by constraints on the spectral index alone, but very modest tightening of the constraint on $n_{\rm S}$ would make this conclusion secure. By contrast, the BICEP2 results act strongly against the curvaton-dominated regime throughout its parameter space \footnote{Though see Ref.~\cite{Sloth:2014sga} for a means by which a curvaton model with a large and negative running spectral index can alleviate the tension between BICEP2 and Planck.}.
 
\section{The inflating curvaton}

We have so far assumed that the curvaton field does not lead to a second phase of inflation after the inflaton has decayed. This was enforced by choosing $\sigma_*\ll \mpl$, since $\sigma_*\gtrsim \mpl/\sqrt{4\pi}$ is required in order to drive a second period of inflation. This condition does not depend on the mass of the curvaton. However, a small vev is not always a requirement to call $\sigma$ a curvaton. 

If the curvaton mass is much smaller than the inflaton mass, then assumption 1 may hold even if the curvaton vev is as large as the inflaton's. This leads to the inflating curvaton scenario, in which the inflaton drives inflation for $N_1$ $e$-folds, then oscillates and decays. The curvaton is still frozen high in its potential, and so it leads to a second phase of inflation for $N_2$ $e$-folds before it also decays. In this scenario, $r_{\rm dec}$ is always unity. We denote the amount of inflation corresponding to the pivot scale of observables by $N_*=N_1+N_2$. 

Any modes which reenter the horizon during the gap between the two inflationary periods, and then re-exit during the second inflationary period, will have an strongly oscillatory pattern imprinted \cite{Mukhanov:1990me}. We will require that all such modes are on smaller scales than we can observationally probe. We hence require $N_1$ to be at least 10, and potentially significantly more if a large range of scales re-enter the horizon during the break between inflationary periods. The gap between the two periods of inflation will last while the Hubble parameter decreases from $H_{1 \rm end}\simeq m_\phi$ until the curvaton's energy density becomes dominant, $H_{2, \rm start}\simeq m_{\sigma}\sigma_*/m_{\rm Pl}$. The range of scales re-entering during this break also depends on whether the background energy density is dominated by radiation or matter. Regardless of how many scales re-enter the horizon during this gap, the total number of $e$-foldings is still required to be roughly between 50 and 60. This is because the total number of $e$-foldings corresponding to a given comoving scale depends only on the energy scale of the first inflationary period and how long the matter and radiation dominated epochs last, and not on the order in which they take place, see Sec.~\ref{nefold} and Ref.~\cite{LiddleLeach}. 

Following the results of Langlois and Vernizzi \cite{Langlois:2004nn} and Vernizzi and Wands \cite{Vernizzi:2006ve}, one finds  
\begin{eqnarray}
\epsilon_*&\simeq&\epsilon_*^\phi\simeq \frac{\mpl^2}{4\pi\phi_*^2}\simeq \frac{1}{2 N_1}, \\ \label{Ninfcurv}
\frac{\partial N}{\partial\phi}&=&\frac{4\pi\phi_*}{\mpl^2}=\frac{\sqrt{8\pi N_1}}{m_{\rm Pl}}, \\
\label{Ninfcurv2}
\frac{\partial N}{\partial \sigma}&=&\frac{4\pi\sigma_*}{\mpl^2}\simeq \frac{\sqrt{8\pi N_2}}{m_{\rm Pl}}, \\
\label{Pratio}\frac{P_{\zeta}^\sigma}{P_{\zeta}^{\phi}}&=&\left(\frac{\sigma_*}{\phi_*}\right)^2,\\ r&=&\frac{4\mpl^2}{\pi\left(\phi_*^2+\sigma_*^2\right)}=\frac{8}{N_*}=r_{\rm single},\\
n_{\rm S}-1&=& -2\epsilon_*\left(1+ \frac{m_\phi^2}{m_{\rm single}^2}\right)\simeq -\frac{1}{N_1} -\frac{1}{N_*} \nonumber\\
& \leq & n_{\rm S,single} -1,
\end{eqnarray}
and $\fnl$ is small \cite{ICTY, Enomoto:2012uy}.
Since the $\delta N$ coefficients, Eqs.~(\ref{Ninfcurv}) and (\ref{Ninfcurv2}), are the same as in two-field quadratic inflation \cite{Vernizzi:2006ve}, the observational predictions are the same and match those of Nflation \cite{DKMW,AlaLyth,Kim:2006ys}. We see from Eq.~(\ref{Pratio}) that the perturbations from the curvaton could be dominant if $N_2$ is big enough, though not by a wide margin. These results are valid for $\sigma_*\gtrsim \mpl/2$, while for $\sigma_*\lesssim \mpl/10$ the standard curvaton results are valid \cite{Langlois:2004nn}.  These limits barely depend on the curvaton mass, provided that it is much less than the inflaton's mass. The intermediate regime was investigated numerically in Ref.~\cite{Langlois:2004nn}.

The conclusion is that the inflating curvaton predicts the same tensor-to-scalar ratio as single-field inflation, and has a redder spectral index and negligible non-Gaussianity. This is in agreement with the predictions of Nflation \cite{AlaLyth,Kim:2006ys,Kim:2007bc}, and hence may be considered as a special, two-field, case of that scenario. It is also the same as the predictions of inflation driven by two quadratic fields which decay at the same time \cite{Vernizzi:2006ve,tarrant,elliston_orani_mulryne}. Finally, note that if one took an equal prior range for $\phi_*$ and $\sigma_*$ then the inflating curvaton would be much more common than the standard curvaton scenario, since it can occur for a much larger range of initial $\sigma_*$ values. 

\section{Conclusions}

\begin{figure}[t!]
\includegraphics[width=0.5 \textwidth]{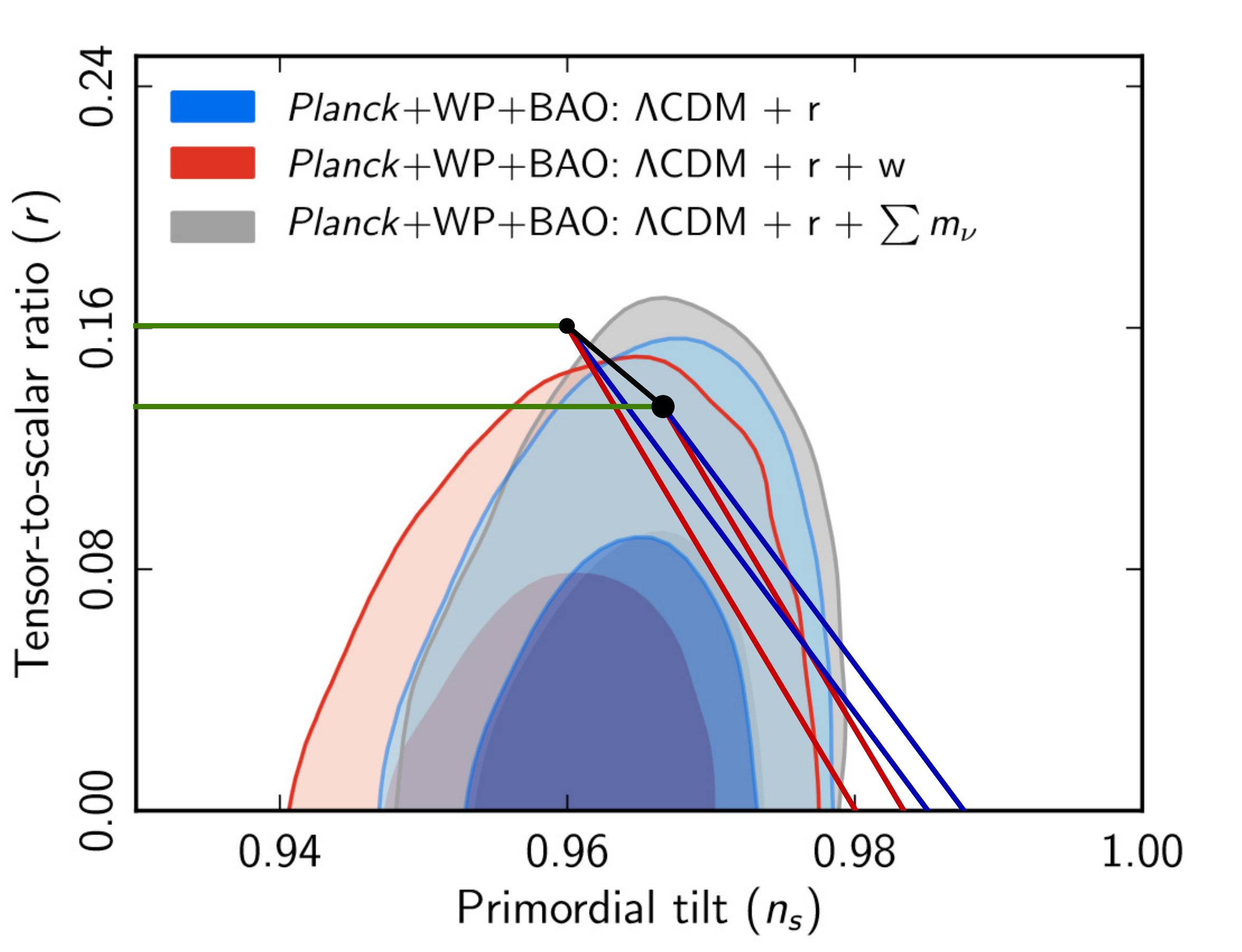}
\caption{Region occupied by curvaton models, allowing $N_*$ to vary between 50 and 60 in all cases. The area between the red lines is the region covered by the usual curvaton scenario when $m_\sigma\ll m_\phi$. The blue lines set $m_\sigma=m_\phi/2$ to show the effect of a massive curvaton. The horizontal green lines are the inflating curvaton regime, which mimics the predictions of Nflation.}
\label{ns-r} 
\end{figure}

In Fig.~\ref{ns-r} we show the locations occupied by curvaton models in the $n_{\rm S}$--$r$ plane in all the regimes we have explored. The simplest curvaton model is in some tension with the {\it Planck} data for all parameter values, primarily due to the observed redness of the spectral index. However the model is not ruled out by this data, and the observational statistical errors are now small enough that any systematic shifts in the spectral index constraints are now important, see e.g.~Ref.~\cite{Spergel:2013rxa}.
Despite the stringent constraint on local non-Gaussianity that the deviations from Gaussianity of curvature perturbation must be less than one part in a thousand, this does not strongly constrain the curvaton scenario. In the curvaton limit, which maximises $\fnl$, the constraint requires that the fraction of the curvaton's energy density at the decay time must satisfy $r_{\rm dec}>0.15$ at $95\%$ confidence \cite{Ade:2013ydc}. This constraint is weakened if the inflaton perturbations are not negligible, $m_{\phi}\simeq m_{\rm single}$.

By contrast, the new BICEP2 results indicating \mbox{$r \gtrsim 0.1$} will, if confirmed, convincingly rule out the pure curvaton limit. They require that the energy scale of inflation is similar to that of quadratic inflation, which requires $m_\phi\sim m_{\rm single}$ and hence that the inflaton perturbations must be comparable to or dominant over the curvaton perturbations. A  significant suppression of $r$ in the curvaton limit is generic for all curvaton models, suggesting that this result has ruled out the curvaton limit (i.e.~the original curvaton scenario assumption \ref{curvaton-limit}) regardless of the choice of inflaton and curvaton potentials.

\begin{acknowledgments}
C.B.~was supported by a Royal Society University Research Fellowship, M.C.\ by EU FP7 grant PIIF-GA-2011-300606, and A.R.L.\  by the Science and Technology
Facilities Council [grant number ST/K006606/1]. Research at Perimeter Institute is supported by the Government of Canada through Industry Canada and by the Province of Ontario through the Ministry of Research and Innovation. We thank Takeshi Kobayashi and David Wands for discussions.
\end{acknowledgments}



\begin{thebibliography}{99}

\bibitem{bicep2}
 BICEP2 Collaboration I (P. A. R. Ade et al.)
 arXiv:1403.3985 

\bibitem{planckI} 
  Planck Collaboration I (P. A. R. Ade et al.)
  arXiv:1303.5062.

\bibitem{planckXXII} 
  Planck Collaboration XXII (P. A. R. Ade et al.),
  arXiv:1303.5082.

\bibitem{seminal} K. Enqvist and M. S. Sloth, Nucl. Phys. B {\bf 626}, 395 (2002) [hep-ph/0109214]; D. H. Lyth and D. Wands, Phys. Lett. B {\bf 524}, 5 (2002) [hep-ph/0110002]; T. Moroi and T. Takahashi, Phys. Lett. B {\bf 522}, 215 (2001) [hep-ph/0110096].

\bibitem{Enq-Tak} K. Enqvist and T. Takahashi, JCAP {\bf 1310}, 034 (2013) [arXiv:1306.5958].

\bibitem{kobayashi} T. Kobayashi, F. Takahashi, T. Takahashi, and M. Yamaguchi, JCAP {\bf 1310}, 042 (2013) [arXiv:1303.6255]. 

\bibitem{ellis} J. Ellis, M. Fairbairn and M. Sueiro, arXiv:1312.1353.

\bibitem{higgs} K. Enqvist, R. Lerner, and T. Takahashi, arXiv: 1310.1374.

\bibitem{tarrant} J. Meyers and E. R. M. Tarrant, arXiv:1311.3972.

\bibitem{BL} N. Bartolo and A. R. Liddle, Phys. Rev. D {\bf 65}, 121301 (2002) [astro-ph/0203076].

\bibitem{CLM} M. Cort\^{e}s, A. R. Liddle, and P. Mukherjee, Phys. Rev. D {\bf 75}, 083520 (2007) [astro-ph/0702170].

\bibitem{Langlois:2004nn} 
  D.~Langlois and F.~Vernizzi,
  Phys.\ Rev.\ D {\bf 70}, 063522 (2004)
  [astro-ph/0403258].

\bibitem{Dimopoulos:2011gb} 
  K.~Dimopoulos, K.~Kohri, D.~H.~Lyth, and T.~Matsuda,
  JCAP {\bf 1203}, 022 (2012)
  [arXiv:1110.2951].

\bibitem{moroi_takahashi_toyoda}
T.~Moroi, T.~Takahashi, Y.~Toyoda,
Phys.\ Rev.\  {\bf D72}, 023502 (2005).
[hep-ph/0501007].

\bibitem{Senoguz:2012iz} 
  V.~N.~Senoguz,
  JCAP {\bf 1210}, 015 (2012)
  [arXiv:1206.4944 [hep-ph]].

\bibitem{WMAP9} C. L. Bennett et al.\ (WMAP collaboration), Astrophys. J. Supp. {\bf 208}, 20 (2013) [arXiv:1212.5225].

\bibitem{planckXVI} Planck Collaboration XVI (P. A. R. Ade et al.), arXiv:1303.5076.

\bibitem{Lyth:2002my} 
  D.~H.~Lyth, C.~Ungarelli, and D.~Wands,
  Phys.\ Rev.\ D {\bf 67}, 023503 (2003)
  [astro-ph/0208055].

\bibitem{Sasaki:2006kq} M. Sasaki, J. Valiviita, and D. Wands, Phys.\ Rev.\ D {\bf 74}, 103003 (2006) [astro-ph/0607627].

\bibitem{LiddleLeach}
A. R. Liddle and S. M. Leach, Phys. Rev. D {\bf 68}, 103503 (2003) [astro-ph/0305263].

\bibitem{Wands:2002bn} 
  D.~Wands, N.~Bartolo, S.~Matarrese, and A.~Riotto,
  Phys.\ Rev.\ D {\bf 66}, 043520 (2002)
  [astro-ph/0205253].

\bibitem{Byrnes:2010ft} 
  C.~T.~Byrnes, M.~Gerstenlauer, S.~Nurmi, G.~Tasinato, and D.~Wands,
  JCAP {\bf 1010}, 004 (2010)
  [arXiv:1007.4277].

\bibitem{Sefusatti:2009xu} 
  E.~Sefusatti, M.~Liguori, A.~P.~S.~Yadav, M.~G.~Jackson, and E.~Pajer,
  JCAP {\bf 0912}, 022 (2009)
  [arXiv:0906.0232].

\bibitem{Becker:2012je} 
  A.~Becker and D.~Huterer,
  Phys.\ Rev.\ Lett.\  {\bf 109}, 121302 (2012)
  [arXiv:1207.5788].

\bibitem{Byrnes:2011gh} 
  C.~T.~Byrnes, K.~Enqvist, S.~Nurmi, and T.~Takahashi,
  JCAP {\bf 1111}, 011 (2011)
  [arXiv:1108.2708].

\bibitem{Kobayashi:2012ba} 
  T.~Kobayashi and T.~Takahashi,
  JCAP {\bf 1206}, 004 (2012)
  [arXiv:1203.3011].

\bibitem{Ade:2013ydc} 
  Planck Collaboration XXIV (P. A. R. Ade et al.)
  arXiv:1303.5084.

\bibitem{Assadullahi:2007uw} 
  H.~Assadullahi, J.~Valiviita, and D.~Wands,
  Phys.\ Rev.\ D {\bf 76}, 103003 (2007)
  arXiv:0708.0223 [hep-ph].

\bibitem{Sloth:2014sga} 
  M.~S.~Sloth,
  arXiv:1403.8051 [hep-ph].

\bibitem{Mukhanov:1990me} 
  V.~F.~Mukhanov, H.~A.~Feldman and R.~H.~Brandenberger,
  Phys.\ Rept.\  {\bf 215}, 203 (1992).

\bibitem{ICTY}
K.~Ichikawa, T.~Suyama, T.~Takahashi, M.~Yamaguchi,
Phys.\ Rev.\  {\bf D78}, 023513 (2008).
[arXiv:0802.4138 [astro-ph]]. 

\bibitem{Enomoto:2012uy} 
  S.~Enomoto, K.~Kohri and T.~Matsuda,
  Phys.\ Rev.\ D {\bf 87}, 123520 (2013)
  arXiv:1210.7118 [hep-ph].

\bibitem{Vernizzi:2006ve} 
  F.~Vernizzi and D.~Wands,
  JCAP {\bf 0605}, 019 (2006)
  [astro-ph/0603799].

\bibitem{DKMW} S. Dimopoulos, S. Kachru, J. McGreevy, and J. Wacker,
  JCAP {\bf 0808}, 003 (2008) [hep-th/0507205].

\bibitem{AlaLyth} L. Alabidi and D. H. Lyth, JCAP {\bf 0605}, 016 (2006) [astro-ph/0510441].

\bibitem{Kim:2006ys} 
  S.~A.~Kim and A.~R.~Liddle,
  Phys.\ Rev.\ D {\bf 74}, 023513 (2006)
  [astro-ph/0605604].

\bibitem{Kim:2007bc} 
  S.~A.~Kim and A.~R.~Liddle,
  Phys.\ Rev.\ D {\bf 76}, 063515 (2007)
  [arXiv:0707.1982].

\bibitem{elliston_orani_mulryne} J.~Elliston, S.~Orani, D.~J.~Mulryne, arXiv:1402.4800

\bibitem{Spergel:2013rxa} 
  D.~Spergel, R.~Flauger, and R.~Hlozek,
  arXiv:1312.3313.




\end{thebibliography}
\end{document}